\documentclass[letterpaper]{article} 
\usepackage{aaai25}  
\usepackage{times}  
\usepackage{helvet}  
\usepackage{courier}  
\usepackage[hyphens]{url}  
\usepackage{graphicx} 
\usepackage{natbib}  
\usepackage{caption} 
\usepackage{booktabs}
\usepackage{numprint}
\usepackage{algorithm}
\usepackage{algpseudocode}
\npthousandsep{,}

\title{Complete Symmetry Breaking for Finite Models}

\newcommand{\miko}{Mikol\'a\v{s} Janota}
\newcommand{\mike}{Michael Codish}
\newcommand{\joao}{Jo\~ao Jorge Ara\'ujo}
\newcommand{\marek}{Marek Dan\v{c}o}
\author {
 \marek\textsuperscript{\rm 1}, \miko\textsuperscript{\rm 1}, \mike\textsuperscript{\rm 2}, \joao\textsuperscript{\rm 3}
}

\affiliations {
    \textsuperscript{\rm 1}Czech Technical University in Prague, Czechia\\
    \textsuperscript{\rm 2}Ben-Gurion University of the Negev, Beer-Sheva, Israel\\
    \textsuperscript{\rm 3}Department of Mathematics, NOVA FCT, Portugal\\
    dancomar@cvut.cz
}

\usepackage[table]{xcolor}
\newcolumntype{G}{>{\columncolor{lightgray}}c}

\usepackage{amsthm,amsmath,amssymb,amsfonts}
\newtheorem{theorem}{Theorem}
\newtheorem{proposition}[theorem]{Proposition}
\newtheorem{lemma}[theorem]{Lemma}
\newtheorem{observation}[theorem]{Observation}

\theoremstyle{definition}
\newtheorem{definition}[theorem]{Definition}

\newtheorem{example}{Example}

\newcommand{\SRC}{\ast}

\newcommand{\DST}{\diamond}
\newcommand{\src}[2]{{#1}\SRC{#2}}
\newcommand{\dst}[2]{{#1}\DST{#2}}

\usepackage{xspace}

\newcommand{\nauty}{\texttt{nauty}\xspace}

\newcommand{\dmc}{\texttt{d4}\xspace}

\newcommand{\greater}{\text{gt}}
\newcommand{\equal}{\text{eq}}
\newcommand{\cp}{\text{pre}}
\newcommand{\vct}[1]{\text{vec}(#1)}
\newcommand{\vctr}[1]{\text{vec}_{r}(#1)}
\newcommand{\vctd}[1]{\text{vec}_{d}(#1)}
\newcommand{\vctc}[1]{\text{vec}_{c}(#1)}

\begin{document}

\maketitle

\begin{abstract}
This paper introduces a SAT-based technique that calculates a compact and
complete symmetry-break for finite model finding, with the focus on structures
with a single binary operation (magmas). Classes of algebraic structures are
typically described as first-order logic formulas and the concrete algebras are
models of these formulas.  Such models include an enormous number of
isomorphic, i.e.\ symmetric, algebras.

A complete symmetry-break is a formula that has
as models, exactly one canonical representative from each
equivalence class of algebras. Thus, we enable answering questions
about properties of the models so that computation and search are
restricted to the set of canonical representations.

For instance, we can answer the question: How many non-isomorphic
semigroups are there of size $n$?  Such questions can be
answered by counting the satisfying assignments of a SAT formula,
which already filters out non-isomorphic models.
The introduced technique enables us calculating numbers of
algebraic structures not present in the literature and going beyond the
possibilities of pure enumeration approaches.
\end{abstract}

%


\section{Introduction}%
\label{sec:Introduction}

Finite model finding has a longstanding tradition in automated reasoning:
often a user is interested in a model rather than proving a
theorem~\cite{mace2}. Models serve as counterexamples to invalid
conjectures~\cite{blanchette-lpar10} but are also interesting on their own.
Indeed, \citet{semigroups4} enumerates semigroups of order 4 on a computer as
early as 1955. A large body of research exists that tackles finite model
finding, enumeration, and counting, using constraint programming (CP) and SAT
techniques~\cite{AudemardB02,distler2014,ag_groupoid,xlnh,lnh,SEM,mace4,paradoxpaper,choiwah-cubes,suda}.

This paper computes compact and complete symmetry breaking constraints
for a wide range of algebraic structures with a single binary operation, known as magmas. 
Classes of algebraic structures, such as semigroups, are typically described as first
order logic formulas and the concrete algebras are models of these
formulas. These models include an enormous number of isomorphic, i.e.\
symmetric, algebras. For a finite algebra over a domain of size $n$,
each permutation of the $n$ domain elements introduces a symmetry. So
there are a super-exponential number of symmetries.  A complete
symmetry breaking constraint for an algebraic structure is satisfied
exactly for the canonical representations of the structure.

Complete symmetry breaks for algebraic structures, as computed in this
paper, are not found in the literature. Applying these enables to
answer, via computation, questions about properties of the models so
that computation and search are restricted to the set of canonical
representations.

For instance, we can answer the question: How many non-isomorphic semigroups
of size $n$ are there? --- As long as we can calculate the complete
symmetry-break. Questions of this type can be answered by counting the
satisfying assignments of a SAT formula encoding both the property of the
algebraic class and the complete symmetry break. Like this, it already filters
out non-isomorphic models. This enables us to use
out-of-the-box \emph{model counters}~\cite{mc} to calculate the non-isomorphic
structures \emph{without} enumerating them.

Our approach can be seen as ``constrain and generate'', whereas pure
enumeration is ``generate and prune''.  Applying our approach enables
us to calculate numbers of algebraic structures not present in the
literature and that go beyond the possibilities of pure enumeration
approaches.

We mention also partial symmetry breaks. These are constraints which
are satisfied by at least one element from each equivalence class of
objects. Partial symmetry breaks are typically smaller in size than
complete breaks but admit redundant (isomorphic) solutions.

The symmetry breaks we compute are lex-leader symmetry breaks. Namely,
from each class of isomorphic (or symmetric) objects, we select a
canonical representative which is the minimal object, with respect to
a lexicographic ordering, in its class.
So, in theory, a complete lex-leader symmetry break is obtained by
introducing a constraint that imposes that an algebraic structure must
not be larger than any of its permutations. A symmetry break defined
in this way involves a super-exponential number of constraints and is
too large to be of practical use.

In this paper we show that symmetry breaks can be often more compact in
practice. However, we also show that unless \texttt{P=NP}, we do not expect to
find a symmetry break for unrestricted algebraic structures of polynomial
size.

This paper adapts the approach described in~\citet{codish} applied to
compute compact and complete symmetry breaks for graph search
problems. In fact, as graphs can be posed as an algebraic structure,
our approach is a generalization of this previous work.
In the case of graphs, \citet{codish} compute a complete symmetry
breaking constraint for order 10 graph search problems consisting of
only \numprint{7853} lexicographic order constraints instead of all
$10!=\numprint{3628800}$ constraints. This is our motivation in the context of
the current paper.
In our context, for example, we compute a complete symmetry break for
AG-groupoids of size 7 that involves 240 lexicographic order constraints
instead of $7! =\numprint{5040}$. Applying this symmetry break
enables counting of all AG-groupoids of size 7, a number which
is not found in current literature.

Reasoning about algebraic structures is significantly harder than about graphs,
algebras correspond to matrices where the contents and row-column indices
interact (one may for example have an axiom $x*x=x$). For binary operations,
the search-space is bounded by $n^{n^2}$, which gets unwieldy as quickly as for
$n=5$ ($\approx$ 3e+17), which is why $n=6$ is already a challenge for many
algebra classes.

\vspace{2pt}
\noindent To summarize the main contributions of this paper:
\begin{enumerate}
    \item We generalize the approach described by~\citet{codish} to apply to
      arbitrary classes of algebraic structures described as first order
      logic formulas.
    \item We compute compact complete symmetry breaks for a variety of
      algebraic structures.
    \item We apply the compact complete symmetry breaks that we compute to
        count non-isomorphic structures of several representative classes of
        algebras obtaining new results.
    \item We give novel theoretical insights in the complexity of
        complete and partial symmetry breaks for
        finite models.
\end{enumerate}


\section{Preliminaries}\label{sec:preliminaries}

In this paper we study finite mathematical structures with a
single binary operation, known as \emph{magmas} (aka \emph{groupoids}).
For example, finite groups or
semigroups are magmas. Magmas are denoted by a pair $(D,\circ)$ where
$D$ is the domain and $\circ$ a binary operation on $D$.  We rely on
the well-established term of isomorphism and isomorphic copy.

\begin{definition}[isomorphism]\label{def:isomorphism}
  A bijection $f:D_1\rightarrow D_2$ is an isomorphism from a magma $(D_1,\SRC)$ to
  $(D_2,\DST)$ if $f(\src{a}{b})=\dst{f(a)}{f(b)}$, for all $a,b\in D_1$. Two
  magmas are \emph{isomorphic} iff there exists at least one isomorphism between them.
\end{definition}

\begin{definition}[isomorphic copy]\label{def:copy}
  Consider a magma $(D_1,\SRC)$ and a bijection $f:D_1\rightarrow D_2$
  then the \emph{isomorphic copy} $(D_2,\DST)_f$ is defined as
  $\dst{a}{b}=f(\src{f^{-1}(a)}{f^{-1}(b)})$.
  For a magma $A$ we write $f(A)$ for its isomorphic copy.
\end{definition}

Throughout the paper, we consider magmas on a finite domain
$D=\{0,\ldots,n-1\}$ for $n\in\mathbb{N}^+$, where $n$ is denoted as
the \emph{size of the magma}. Then, we are only concerned with
isomorphisms between magmas on the same domain $D$, which means that
any such isomorphism is a permutation $\pi$ on $D$, i.e., an element of the
relevant symmetric group, which we will denote as $S_n$.

A finite magma can be naturally represented as a two-dimensional
multiplication table, which enables us to order them
lexicographically, comparing their vectorization. The table may be
vectorized in an arbitrary fashion, as long as this vectorization is
fixed.

\begin{definition}[magma vectorization]
  Let $A$ be a magma. Then $\vct{A}$ is a fixed vectorization of the
  two-dimensional multiplication table corresponding to $A$.
\end{definition}

Throughout the paper we consider three alternative vectorizations of
magmas. Let $\vctr{A}$ denote the ``row-by-row'' concatenation of
the rows of $A$ left-to-right, top-down, i.e., the way people are used
to read and write in Western civilization.
Let $\vctd{A}$ denote the ``diagonal first'' vectorization in which
the elements of the diagonal occur first, followed by the row-by-row
vectorization, skipping the diagonal elements.
Let $\vctc{A}$ denote the ``concentric'' vectorization  where
cells are ordered by the sum of row and column indices and ties are
broken by row-by-row ordering.
So for example, \text{xor},
i.e.~{\tiny$\begin{array}{c}01\\10\end{array}$}, is vectorized as
$0,1,1,0$ in  $\vctr{\cdot}$,$\vctc{\cdot}$ and as $0,0,1,1$ in $\vctd{\cdot}$.
For $x+y \mod 3$,
i.e.~{\tiny$\begin{array}{c}
012\\
120\\
201
\end{array}$},
$\vctr{\cdot}$ gives $0,1,2,1,2,0,2,0,1$,
$\vctc{\cdot}$ gives $0,1,1,2,2,0,2,0,1$, and
$\vctd{\cdot}$ gives $0,2,1,1,2,1,0,2,0$.
These vectorizations are mutually incompatible but each individually impose
a total order on algebras via the lexicographic ordering of the
corresponding vectors.

\begin{definition}[$\preceq$]
  Assume a vectorization $\vct{\cdot}$ for magmas of size $n$.
  Define a total order $\preceq_{\vct{\cdot}}$ by defining $A\preceq_{\vct{\cdot}} B$ true
  iff $\vct{A}$ is lexicographically smaller or equal than $\vct{B}$.
  Note that $\vct{A}$ is a vector of length $n^2$.
  We usually omit the subscript and write  $\preceq$ when $\vct{\cdot}$ is
  clear from the context.
\end{definition}

\begin{definition}[lex-leader]
  We say that a magma $A=(D,\SRC)$ is a \emph{lex-leader} with respect
  to a given vectorization ${\vct{\cdot}}$, iff
  for all  magmas $B$ isomorphic to $A$, it holds that $A\preceq_{\vct{\cdot}} B$.
\end{definition}


\section{Canonizing Sets Preliminaries}\label{sec:gr:preliminaries}

Let $\mathcal{M}_n$ denote the set of all order $n$ algebraic
structures of a selected type (e.g.\ quasigroups).

The following is adapted from~\cite{codish} where the definition and
algorithm are stated for simple graphs.

\begin{definition}[canonicity]\label{def:canonizing}

  Let $A\in\mathcal{M}_n$, $\Pi\subseteq S_n$, and denote the
  predicate
  $\min_\Pi(A) = \bigwedge_{\pi\in\Pi} A \preceq \pi(A)$.
  We say that $A$ is canonical if $\min_{S_n}(A)$. We say
  that the set $\Pi$ is \emph{canonizing} if
  $%
  \forall_{A\in \mathcal{M}_n}.\,\min_\Pi(A) \leftrightarrow \min_{S_n}(A)
  $.
\end{definition}

In a nutshell, a canonizing set $\Pi$ is a means of \emph{quantifier
elimination}~\cite{bradley-manna07} by replacing $\forall\pi\in S_n$ by
$\bigwedge_{\pi\in\Pi}$ with the aim of obtaining $\Pi$ smaller than $n!$.
Algorithm~\ref{CanSetAlgorithm} gradually augments $\Pi$ through
counter-examples until it becomes canonizing.
It starts with some initial set of permutations $\Pi$
(for simplicity, assume that $\Pi=\emptyset$). Then, incrementally
apply the step specified in lines~\ref{line:while}--\ref{line:bla} of
Algorithm~\ref{CanSetAlgorithm}, as long as the stated condition
holds.
\begin{algorithm}
\begin{algorithmic}[1]
\State \textbf{Init}: $\Pi=\emptyset$
\While{ $\exists A\in \mathcal{M}_n$ $\exists \pi\in S_n$ s.t.\ 
        $\min_\Pi(A)$  and $\pi(A)\prec A$}\label{line:while}
\State  $\Pi=\Pi\cup\{\pi\}$\label{line:bla}
\EndWhile
\State \textbf{return} $\Pi$
\end{algorithmic}
\caption{Compute Canonizing Set}%
\label{CanSetAlgorithm}
\end{algorithm}

\begin{lemma}[\citet{codish}]
  Algorithm~\ref{CanSetAlgorithm} terminates and returns a canonizing
  set $\Pi$.
\end{lemma}


We say that a canonizing set $\Pi$ of permutations is \emph{redundant} if for
some $\pi\in\Pi$ the set $\Pi\setminus\{\pi\}$ is also canonizing.
Algorithm~\ref{CanSetAlgorithm} may compute a redundant set. For
example, if a permutation added at some point becomes redundant in
view of permutations added later. An algorithm \emph{Reduce Algorithm}
is straightforward to implement and we denote it as \textbf{Algorithm 2}. It
iterates on the elements of a canonizing set to remove redundant
permutations (similarly to the iterative algorithm for minimally unsatisfiable set or monotone predicates in general~\cite{humus,monotone}).


\section{Canonizing Sets for Algebras}\label{sec:algebra}

\subsection{Encodings}%
\label{sub:Encodings}

To apply Algorithm~\ref{CanSetAlgorithm}
it is needed to encode into SAT that we are looking for an
interpretation $A$ that is a model of some given first order logic
axioms with domain size $n$.  This is done by standard means,
cf.~\cite{mace2,paradoxpaper}.
Notably, for each triple of domain elements $(d_1, d_2, d_3)$ a
propositional variable $x_{d_1,d_2,d_3}$ is introduced, representing
the fact that $d_1\ast d_2=d_3$ in the model $A$; for ease of
notation, we write $A[d_1,d_2]\approx d_3$ to refer to this variable.
These variables encode the value of $x*y$ in \emph{one-hot} encoding,
expressed as a cardinality constraint, encoded into CNF by standard
means~\cite{cardinality_handbook}, i.e.\
$1 = \Sigma_{d\in D} A[d_1,d_2]\approx d$, for $d_1,d_2\in D$.

First, we present the encoding of the while loop condition at
line~\ref{line:while} in Algorithm~\ref{CanSetAlgorithm}.
The condition has four parts: $\exists A\in \mathcal{M}_n$,
$\exists \pi\in S_n$, $\min_\Pi(A)$, and $\pi(A)\prec A$. The encoding
is a conjunction:
$\varphi_1 \wedge \varphi_2 \wedge \varphi_3 \wedge \varphi_4$ of four
corresponding propositional formulas.
The overall complexity of the encoding is $O(n^7)$, however, simplifications are used for fixed permutations see~\citet[Sec~3.2.2]{diplomka}.

\paragraph{The first conjunct}  comprises the constraints imposed by
the FOL specification which specifies $\mathcal{M}_n$.
\paragraph{The second conjunct} models that $\pi$ is an (unknown)
permutation. To this end, we introduce propositional variables
$\pi[d_1]\approx d_2$ for $d_1,d_2\in D$, to represent the permutation
$\pi$, under the constraints that they indeed behave like a
permutation:
$1=\Sigma_{d\in D}\pi[d]\approx d'=\Sigma_{d\in D}\pi[d']\approx d$,
for $d\in D$.

\paragraph{The fourth conjunct} is the encoding of the constraint $\pi(A)\prec A$
where both $A$ and $\pi$ are unknown (existentially quantified).  This
is the crux of the condition (we will come back to the third
conjunct later).
\citet{codish} show how to encode this constraint for graphs but for
algebras, this encoding is much more involved because the matrix
contains values and permutations apply also to these
values.

To reason about $\pi(A)$, it is  useful to realize that if $\pi(A)$ has the
value $d$ in cell $r,c$, the multiplication table $A$ will have the value
$\pi^{-1}(d)$ in cell $\pi^{-1}(r), \pi^{-1}(c)$ ($\pi$ is effectively a
renaming). Since the SAT encoding does not enable us to treat the permutation
$\pi$ as a first-order citizen, we will have to go through all possible
combinations of pre-images.
To simplify the presentation, we introduce an auxiliary subformula $\cp$,
expressing that $r',c',d'$ are pre-images of $r,c,d$ under $\pi$, respectively,
and the value of $A$ is $d'$ at the cell $r',c'$.

\[
  \begin{array}{l}
    \cp(r',r,c',c,d',d) \equiv \\
  \pi[r']\approx r\land \pi[c']\approx c\land \pi[d']\approx d\land A[r',c']\approx d'
  \end{array}
\]

Note that if $\cp(r',r,c',c,d',d)$ is true then $\pi(A)$ has the value $d$ at
position $r, c$ because, effectively, $\pi$ replaces the primed values with
their non-primed version.

We consider some fixed vectorization of the multiplication table, which is
represented as a sequence of row-column index pairs,
$(r_1,c_1),\dots,(r_{n^2},c_{n^2})$. Now we can define auxiliary propositional
variables that express that the value of $A$ in cell $r_i,c_i$ is greater/equal than
$\pi(A)$ in the same cell by going through all combinations of pre-images and
values.

\begin{align*}
    &\greater^d_i \equiv A[r_i, c_i] \approx d \Rightarrow \bigvee_{\substack{r',c',m',m \in D \\ m<d}} \cp(r',r_i,c',c_i,m',m)\\
    &\greater_i \equiv \bigwedge\nolimits_{d\in D}\greater^d_i\\
    &\equal^d_i \equiv A[r_i, c_i] \approx d \Rightarrow \bigvee_{\substack{r',c',d'\in D}} \cp(r',r_i,c',c_i,d',d)\\
    &\equal_i \equiv \bigwedge\nolimits_{d\in D}\equal^d_i\\
\end{align*}

Finally, we introduce auxiliary propositional variables \(r_1, \dots, r_{n^2 - 1}\)  to reason about suffixes of $A$,
i.e.\ if $r_i$ is true than $\vct{A}<\vct{\pi(A)}$ starting at index $i$,
which lets us express the inequality between $\pi(A)$ and $A$.

\begin{align}\label{pi(A)<Ab}
  \pi(A) \prec A \:\equiv \: & ( \greater_1 \lor ( \equal_1 \: \land \: r_1))\land{}  \nonumber\\
                             &\bigwedge\nolimits_{2\;\leq\;i\;\leq\;n^2-1}
                             r_{i-1} \Rightarrow ( \greater_i
                             \lor \: ( \equal_i \: \land \: r_i )) \nonumber\\
                             & \land \: ( r_{n^2 - 1} \Rightarrow \greater_{n^2}) \nonumber\\
\end{align}

\paragraph{The third conjunct} is the encoding of the constraint
$\min_\Pi(A) = \bigwedge_{\pi\in\Pi}{A \preceq \pi(A)}$. Here,
$\Pi$ is a set of known permutations and $A$ is the unknown matrix
from the other conjuncts. We skip the details of the encoding. For
each known $\pi$ in $\Pi$ we have a conjunct $A \preceq \pi$ which
is the negation of the encoding from the fourth conjunct, equation~\eqref{pi(A)<Ab}, and
simpler as the permutation is known.


\subsection{Complexity Insights}%
\label{sub:bounds}

%
For graphs, complexity of lex-leader has been studied.
\citet{Babai1983} show that lex-leader adjacency matrix construction is NP-hard.
\citet{crawford-kr96} show that deciding whether a given \emph{incidence} matrix is a lex-leader is NP-complete.
Here we show that lex-leaders of graphs can be ``simulated'' in algebras.
For a given undirected graph $G = (V, E)$, without self-loops, define the FOL theory $F_G$:
\begin{align}
   & \forall x.c*x=x*c=c &\text{ (force $c$ to 0)}\label{eq:idem1}\\
   & \forall x.\, x*x=c&\text{\ (nothing else is idempotent)}\label{eq:idem2}\\
   & \bigwedge\nolimits_{a,b\in V\cup\{c\}} a\neq b&\text{\ (distinct vertices and 0)}\\
   & v_1*v_2 \neq c &\{v_1,v_2\}\in E\,\text{(non-0 on edges)}\label{eq:nonzero} \\
   & v_1*v_2 = c &\{v_1,v_2\}\notin E\,\text{(0 on non-edges)}\label{eq:zero}
\end{align}

\begin{observation}
  Any lex-leader under the row-by-row ordering of a model with the domain $\{0,\dots |V|\}$ of $F_G$ corresponds to a lex-leader of 
  the adjacency matrix of $G$.
\end{observation}
\begin{proof}[Proof sketch]
    Since $c$ is the only idempotent and $A$ is a lex-leader, $A(c)=0$,
    from~\eqref{eq:idem1},~\eqref{eq:idem2}. Further, $A$ will only place a
    non-zero value on cells with $A(v_1)*_A A(v_2)$ with $\{v_1,v_2\}\in E$ due
    to~\eqref{eq:nonzero},~\eqref{eq:zero}. Therefore, the sub-table $1..|V|
  \times 1..|V|$ describes the adjacency matrix of $G$, where $A(v_1)*_A A(v_2)\neq 0$
  iff $\{v_1,v_2\}$ is an edge in $G$.
\end{proof}



The observation leads us to believe that deciding whether an algebra is a
lex-leader is NP-hard (but we have not proven it).
On the other, if there existed a canonizing set $\omega$ with polynomially number of
permutations, there would be a polynomial algorithm to decide whether a given
algebra is a lex-leader by testing only the permutations in $\omega$.
Hence, in the general case, we expect super-polynomial lower bounds on the size of canonizing sets.

In contrast, the following example shows the other extreme and that is, if we
focus on a specific class of magmas, set of permutations can be drastically
reduced.

\begin{example}\label{canset:one}
  Consider the FOL unit clause $x\ast y=z\ast w$. For any of its model $A$
  we have $A(x, y) = d$ for all $x, y \in D$ for some fixed $d \in D$.
  For every domain size there exists exactly one isomorphism class and the
  lex-leader multiplication table consists of all zeros (with respect to any ordering of
  cells). There exists a minimal canonizing set for any domain of size $n$
  containing exactly the permutation consisting of a single cycle $(n-1 \; n-2
  \; \dots \; 1 \; 0)$.
\end{example}

\subsection{Partial Symmetry Breaks}%

Partial symmetry breaks for finite models are considered in the literature,
notably \emph{Least number heuristic (LNH)}~\cite{lnh,SEM,xlnh,choiwah-cubes}
is a symmetry break for finite model finding that can be used both dynamically and
statically~\cite{paradoxpaper,suda}. The intuition behind the break is that in a
partially filled table, values that do not yet appear are indistinguishable and
therefore the smallest one can be used to represent all of them.
So for example, the break will restrict the values of the cell $(0,0)$ to $\{0,1\}$ because
$1$ is the smallest of all the yet unseen values
($0$ is taken to be seen already because is in the index of the cell).

Any set of permutations $\omega\subseteq S_n$ gives us a partial
symmetric break for an algebra $A$ by constructing the constraint
$\bigwedge_{\pi\in\omega} A\preceq\pi(A)$. A transposition is a
permutation which interchanges two elements and leaves all others
fixed. We show that LNH is strictly weaker than breaking by
transpositions only, i.e.\ by considering only $\binom{n}{2}$
elements of the symmetric group $S_n$. We note that breaking 
symmetries with transpositions is a popular technique for graph
search problems and is considered in~\cite{Codish2018}.

\begin{definition}
  We say that a magma $A$ is \emph{minimal with respect to the set of all
  transpositions} iff for all transpositions $\tau\in S_n$ we have $A
  \preceq \tau(A)$.
\end{definition}

\begin{proposition}\label{transp:obs}
  Any magma minimal with respect to the set of all transpositions also satisfies the LNH\@.
\end{proposition}
\begin{proof}
  Suppose we have a magma $A$ and the smallest~$i$ such that there exists a
  cell $A(r_i, c_i)$ with $A(r_i, c_i) = d$ where $d$ is larger than the
  maximal designated number $m$. Then this model does not satisfy the LNH\@.
  Moreover, for this model also holds $\tau(A) \prec A$ for $\tau = (m \;
  d)$, therefore $A$ also is not minimal with respect to the set of all
  transpositions.
\end{proof}

\begin{example}\label{transp:exa}
  Suppose $A = \langle D, \diamond \rangle$ is a magma of order 4 with $3
  \ast^A 3 = 2$ and $ x\ast^A y=0$ for all other $x,y \in D$. Then $A$ is
  permitted by the LNH\@. However, $A$ is not minimal with respect to the set of
  all transpositions since for $\tau = (1 \; 2)$ and $\tau(A) = \langle D,
  \circ \rangle$ we have $\tau(A) \prec A$. $A$ and $\tau(A)$ are depicted
  in Figure~\ref{transposition}.
\end{example}

\begin{figure}
\centering
  \[
  \begin{array}{c| *4{c}}
    \diamond & 0 & 1 & 2 & 3 \\
    \hline
    0 & 0 & 0 & 0 & 0 \\
    1 & 0 & 0 & 0 & 0 \\
    2 & 0 & 0 & 0 & 0 \\
    3 & 0 & 0 & 0 & 2
  \end{array}
  \hspace{1,5cm}
  \begin{array}{c| *4{c}}
    \circ & 0 & 1 & 2 & 3  \\
    \hline
    0 & 0 & 0 & 0 & 0 \\
    1 & 0 & 0 & 0 & 0 \\
    2 & 0 & 0 & 0 & 0\\
    3 & 0 & 0 & 0 & 1
  \end{array}
  \]
\caption{$A$ that satisfies LNH and $\tau(A)$ with $\tau(A) \prec A$.}%
\label{transposition}
\end{figure}

\begin{observation}\label{transp:weak}
  There is a magma, which is not transposition-minimal  (and therefore not a
  lex-leader) but permissible under LNH\@  (by observing
  Example~\ref{transp:exa}).
\end{observation}



\section{Experiments}
\label{sec:Experiments}
\setlength{\tabcolsep}{1mm}

We implemented Algorithms~\ref{CanSetAlgorithm} and~2 in \texttt{Python3} using the
\texttt{PySAT} package~\cite{pysat} with \texttt{CaDiCaL~1.9.5}~\cite{cadical}
as the back-end SAT solver. The knowledge-compilation-based tool \dmc~\cite{d4}
was used for model counting.
We chose \dmc for its support of \emph{projected} model counting, which is
needed due to auxiliary variables in the encoding.
The experiments were run on a server with four AMD~EPYC~7513 32-Core
Processor@2.6GHz and with 504~GB of memory. For each problem, we set a memory
limit of 50GB and a timeout of 24 hours.
We use the row-by-row ($\vctr{\cdot}$) and diagonal ($\vctd{\cdot}$)
vectorization as the bases of the lexicographic ordering (see Preliminaries),
the concentric vectorization behaved similarly to the row-by-row one and is not
included in the experiments. For more detailed analysis of the experiments see \citet{diplomka}.



\newcounter{algebrano}
\newcommand{\ralgebra}[1]{\refstepcounter{algebrano}\label{#1}}
\ralgebra{agg}
\ralgebra{cmq}
\ralgebra{grp}
\ralgebra{imz}
\ralgebra{isg}
\ralgebra{ipl}
\ralgebra{loo}
\ralgebra{mag}
\ralgebra{med}
\ralgebra{mon}
\ralgebra{qgr}
\ralgebra{rgr}
\ralgebra{rim}
\ralgebra{sgr}

\begin{table}[t]
\centering
\begin{tabular}{p{2.2cm}p{5.5cm}}
\textbf{Algebra class} & \textbf{Definition in FOL} \\\midrule
A\ref{agg}:AG-groupoid & \((x y) z=(z y) x\).\\\midrule

A\ref{cmq}:Comm.\ quasigroup & Quasigr.\ +\ $xy=yx$.\\\midrule

A\ref{grp}:Group & Monoid\ +\ $xx^\prime = e, x^\prime x = e$\\\midrule

A\ref{imz}:Impl.\ zroupoid &
\begin{tabular}[t]{@{} l@{}}
$(x y) z = (((z e) x) ((y z) e)) e, $ \\
$ \,(e e) e = e$.
\end{tabular} \\\midrule

A\ref{isg}:Inverse \;semigroup &
\begin{tabular}[t]{@{}l@{}}
$(xy)z=x(yz), x=xx'x$,\\
$x''=x, (xx')(yy')=(yy')(xx')$.
\end{tabular} \\\midrule

A\ref{ipl}:Inverse property loop & Loop\ +\ $x^\prime (xy)=y, (yx)x^\prime = y$.\\\midrule


A\ref{loo}:Loop & Quasigr.\ +\ $ex=x,\, xe=x$.\\\midrule

A\ref{mag}:Magma & No requirement.\\\midrule
A\ref{med}:Medial quasigroup & Quasigr.\ +\ $(xy)(wz)=(xw)(yz)$.\\\midrule

A\ref{mon}:Monoid & $x (y z) = (x y) z,  x e = x,  e x = x$.\\\midrule

A\ref{qgr}:Quasigr. & $x y = x z \Rightarrow y = z,  y x = z x \Rightarrow y = z$.\\\midrule

A\ref{rgr}:Rectang.\ groupoid & $(wx = yz) \Rightarrow (wx = wz)$.\\\midrule

A\ref{rim}:Right involutory magma &
\begin{tabular}[t]{@{}l@{}}
$(x y) y = x$.
\end{tabular} \\\midrule
A\ref{sgr}:Semigr. & $x (y z) = (x y) z$. \\
\end{tabular}
\caption{FOL definitions of the used algebraic structures.}\label{tab:algebry}
\end{table}


Several representative algebra classes with domain sizes $n\in 2..10$ are considered---we recall
that the search-space of size $n^{n^2}$ becomes unwieldy already for $n=5$.
The axiomatization  of the algebra classes can be found in Table~\ref{tab:algebry}---for the
sake of succinctness, the infix operator ${}*{}$ is omitted. 
Strictly speaking, some of these classes of algebras are
not magmas because they 
are equipped with ${}'$.
However, our approach is applicable because in all these cases, the unary 
operation ${}'$ is uniquely determined by ${}*{}$.

A group is an associative quasigroup with identity, making
associative magmas (semigroups) and quasigroups the most
studied magmas. Consequently, we mainly focus
on these structures and several of their subclasses, including inverse
semigroups, which is, after groups, the most important and studied class of
semigroups.
AG-groupoids, originally  called \emph{left almost semigroups},
were included in our study, as they were previously tackled by
\citet{ag_groupoid} as an important algebra class.
Zroupoids were introduced as a generalization of two important 
algebras: De Morgan algebras (a generalization of Boolean algebras)
and join-semilattices with zero~\cite{demorgan}.
Numbers for groups, magmas, and semigroups are known for larger values of $n$ due to theoretical insights.

Rectangular groupoids generalize two other algebraic classes:
rectangular semigroups 
and central groupoids, 
cf.~\cite{rectangular_groupoid_zavedenie}.
Right involutory magmas were introduced recently, in the context of
set-theoretic solutions for the Yang-Baxter equation
~\cite{right_involutory_magma_zavedenie}.

\begin{table}[t]
  \centering
  \begin{tabular}{p{2.5em}p{1em}cccc}
    {type} & $n$ &  \# & {time} & {\#models} & {mc-time} \\\midrule
\textbf{A\ref{agg}}  &  7 & \numprint{240}  &  \numprint{1013} & \textbf{\numprint{643460323187}}  &  \numprint{72825}\\    
A\ref{cmq}  &  7 & \numprint{96}  &  \numprint{327} & \numprint{6381}  &  \numprint{186}\\  
A\ref{grp}  &  10 & \numprint{18}  &  \numprint{149} & \numprint{2}  &  \numprint{31}\\  
\textbf{A\ref{imz}}  &  7 & \numprint{173}  &  \numprint{371} & \textbf{\numprint{600767308670}}  &  \numprint{48816}\\  
\textbf{A\ref{imz}} &  6 & \numprint{35}  &  \numprint{22} & \textbf{\numprint{34810736}}  &  \numprint{43}\\  
A\ref{isg}  &  10 & \numprint{317}  &  \numprint{18202} & \numprint{169163}  &  \numprint{10697}\\  
A\ref{ipl} &  10 & \numprint{45}  &  \numprint{312} & \numprint{47}  &  \numprint{87}\\  
A\ref{loo} &  7 & \numprint{78}  &  \numprint{40} & \numprint{23746}  &  \numprint{53}\\  
A\ref{mag}  &  4 & \numprint{23}  &  \numprint{2} & \numprint{178981952}  &  \numprint{145}\\  %
A\ref{med} &  10 & \numprint{29}  &  \numprint{2844} & \numprint{19}  &  \numprint{372}\\  
A\ref{mon}  &  8 & \numprint{218}  &  \numprint{1236} & \numprint{1668997}  &  \numprint{1830}\\  
A\ref{qgr}  &  6 & \numprint{124}  &  \numprint{101} & \numprint{1130531}  &  \numprint{717}\\  
\textbf{A\ref{rgr}}  &  6 & \numprint{158}  &  \numprint{241} & \textbf{\numprint{52574246}}  &  \numprint{6146}\\  
A\ref{rim}  &  6 & \numprint{554}  &  \numprint{411} & \numprint{267954164}  &  \numprint{12070}\\  
A\ref{sgr}  &  7 & \numprint{218}  &  \numprint{872} & \numprint{1627672}  &  \numprint{983}\\  
  \end{tabular}
  \caption{Highest $n$ for which model count was obtained for algebras in Table~\ref{tab:algebry}.
    The column \# is the size of the canonizing set and ``time'' the total time of its calculation.
    The last two columns report on the model counting by \dmc.
    Novel results are highlighted in bold.
  }\label{tab:best}
\end{table}

Table~\ref{tab:best} summarizes the main achieved results. For each algebra class we
list the highest size ($n$) where we were able to calculate the number of
non-isomorphic models. In the case of implication zrupoids (A\ref{imz}), we list both $n=6$
and $n=7$  because none of these are found in the literature. The size of the
canonizing set (column ``\#'') is measured in the number of permutations it
contains.

These results highlight the power of our approach. For
instance~\citet{ag_groupoid} devise a dedicated techniques to count
AG-groupoids (A\ref{agg}) but are only able to get to size~$6$ (where there are already
\numprint{40104513} algebras) but it is still based on enumeration. In contrast,
 our approach is able to count them for $n=7$, which is 4 orders of magnitude
larger and beyond the possibilities of enumeration.

The next values of $n$ were not reached due to either failing to compute a canonizing set
for the corresponding algebra or the model counter exceeding the time limit, with the former 
being the predominant case. Canonizing set computation was slightly more often limited by time than by 
memory (29 to 23), with memory limitations typically arising for $n=10$ due to the complexity of the encoding. 
We computed canonizing sets for magmas up to $n=6$, and for implication 
zroupoids and AG-groupoids up to $n=8$.

As expected, the more structure the algebra class has, the smaller the canonizing
set. An extreme case are groups (A\ref{grp}), where there are only 2 groups of
size $10$ and only 18 permutations are sufficient to break all the symmetries.
In contrast, rectangular groupoids (A\ref{rgr}) are ``loosely defined'', and
quickly lead to a large number of non-isomorphic models as well as the number of
permutations in the canonizing set.

For several of these classes it is possible to verify the counts at \emph{The
On-Line Encyclopedia of Integer Sequences (OEIS)}~\cite{oeis}. 
%
%
We also remark that a closed formula is known for magmas~\cite{harrison1966}.

\begin{figure}[t]
   \centering
   \includegraphics[width=.4\textwidth]{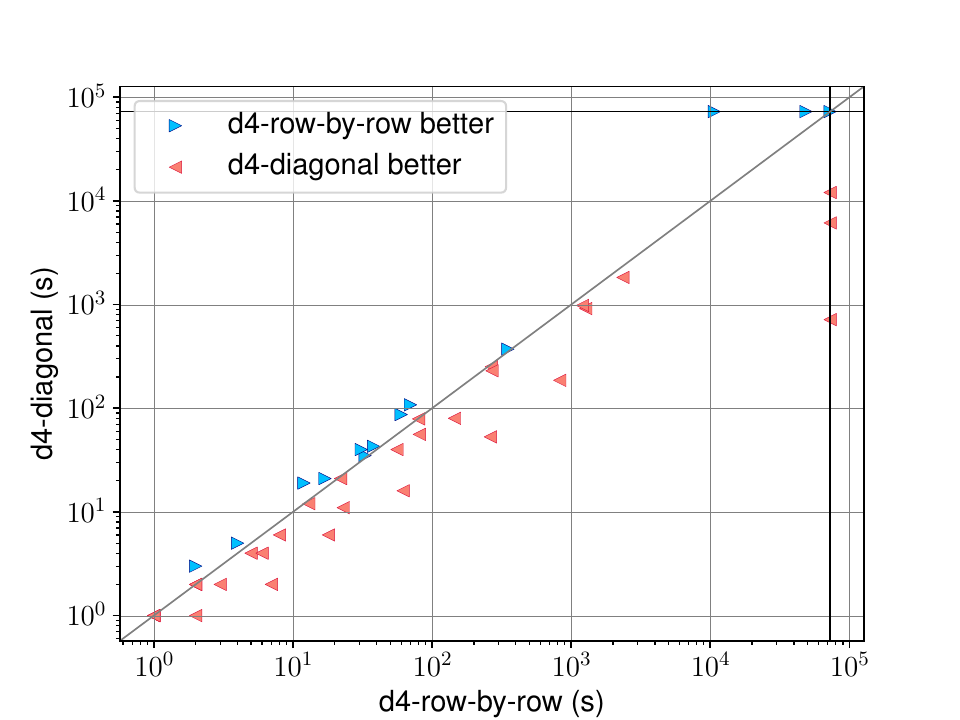}
   \caption{Comparison of the runtime \dmc model counter on row-by-row and diagonal ordering.}\label{fig:d4runtimes}
\end{figure}
\begin{figure}[t]
   \centering
   \includegraphics[width=.4\textwidth]{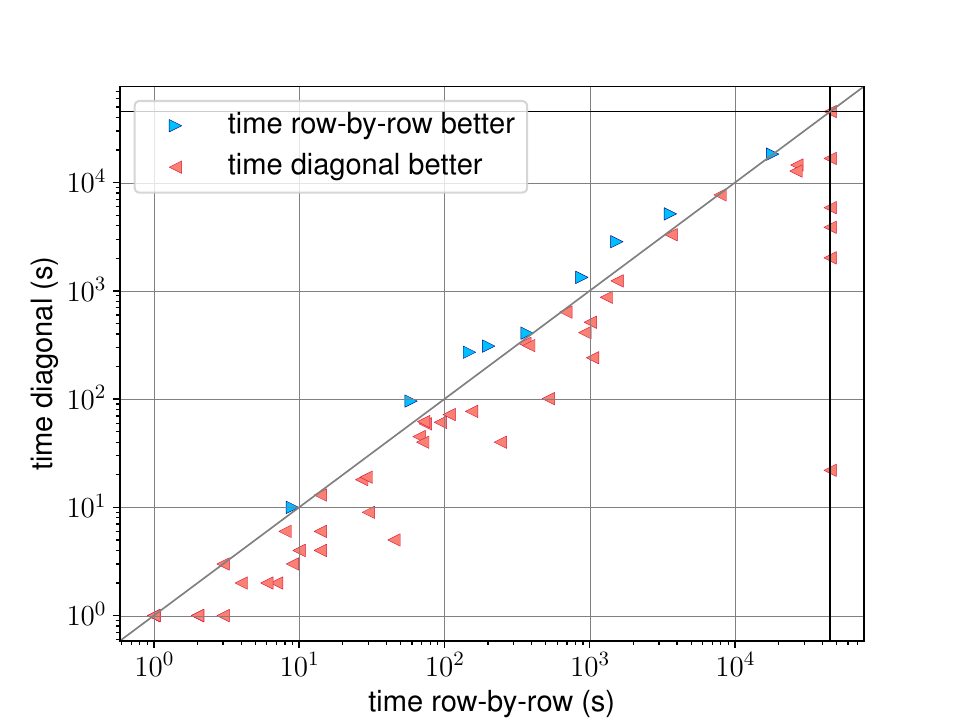}
   \caption{Comparison of the times of canonizing set on row-by-row and diagonal ordering.}\label{fig:scatter:times}
\end{figure}
\begin{figure}[t]
   \centering
   \includegraphics[width=.4\textwidth]{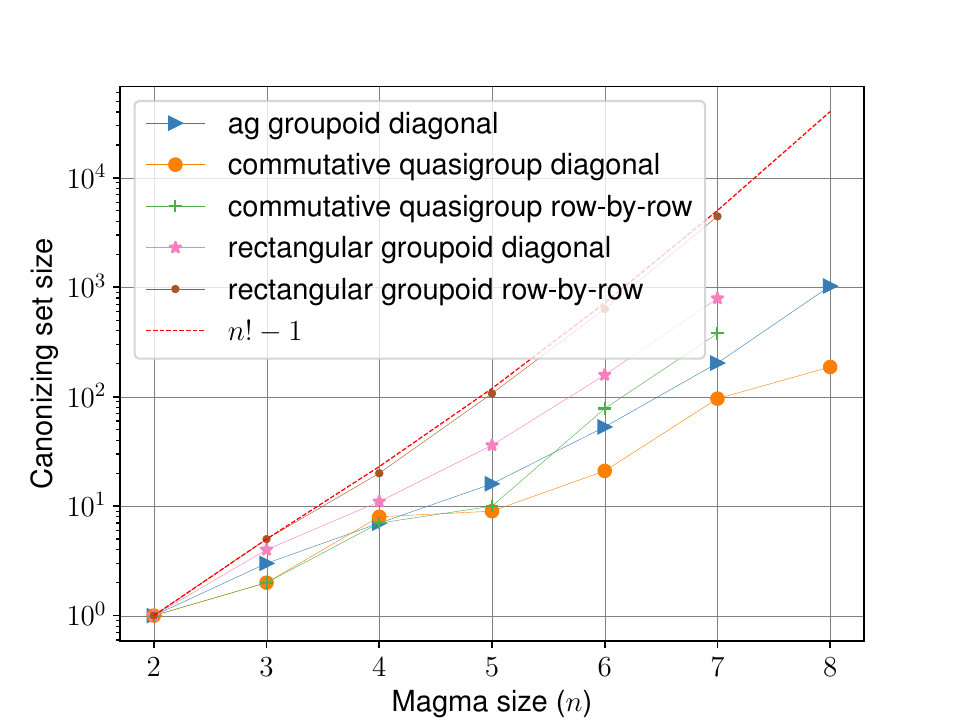}
   \caption{Sizes of canonizing set on selected algebra classes.}\label{fig:plot:sizes}
\end{figure}
\begin{figure}[t]
   \centering
   \includegraphics[width=.4\textwidth]{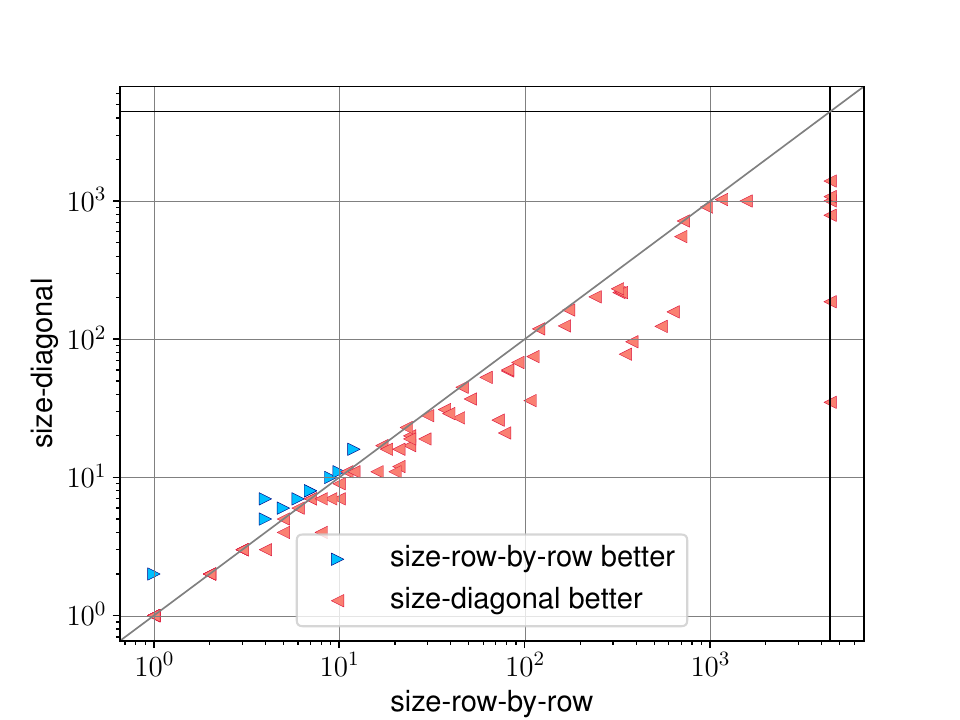}
   \caption{Comparison of the sizes of canonizing set on row-by-row and diagonal ordering.}\label{fig:scatter:sizes}
\end{figure}

Figures~\ref{fig:d4runtimes}--\ref{fig:scatter:sizes} provide further insights into
the experiments. Figure~\ref{fig:d4runtimes} presents the runtime of
the model counter \dmc on all the considered algebras, comparing the two
considered orderings. Figure~\ref{fig:scatter:times} shows the total times of
the calculation of the canonizing set (including the reduction).

Figure~\ref{fig:plot:sizes} shows for selected algebras how the size
of the canonizing set grows with~$n$. Note that the sizes are plotted in log-scale
and form almost a perfect straight line, which is indicative of
exponential growth.  Recall that $n!-1$ is a trivial upper bound on
the canonizing set size and the row-by-row ordering for
rectangular groupoids is very close to it.
In contrast, for AG-groupoids and commutative quasigroup the slope is much
less steep, which enables us to get to $n=8$. For AG-groupoids, the corresponding DIMACS has 239MB\@ with
\numprint{1154} permutations (note that $n!=\numprint{40320}$).

Diagonal ordering gives smaller canonizing sets than the row-by-row
ordering (Figure~\ref{fig:scatter:sizes}); it is the subject of future work to understand why.
This does not always lead to better performance in the model counter; in several
outliers, when the canonizing sets are not too different in size, \dmc runs
out of memory on the diagonal and not in the row-by-row. However, in the majority
cases, the smaller canonizing set provided by the diagonal also leads to faster
times in \dmc (Figure~\ref{fig:d4runtimes}). For example, the new result for
implication zrupoids (A\ref{imz}, $n=7$) and rectangular grupoids (A\ref{rgr}, $n=6$)
were only possible under the diagonal ordering.

\section{Related Work}\label{sec:related}

Finite model finding has a longstanding tradition in automated
reasoning.  Sometimes, a user is interested in a model rather than
proving a theorem~\cite{mace2}. Models serve as counterexamples to
invalid conjectures~\cite{blanchette-lpar10}, which also appear in
formal methods, e.g.\ in Software
Verification~\cite{torlak-tacas07}. Finite models can also be used as
a semantic feature for \emph{lemma selection
  learning}~\cite{urban-ijcar08}.  A number of CP-based methods exists
that enumerate (all) models, cf.~\citet{choiwah-cubes}.
Finite models are also often constructed by dedicated
algorithms anchored in domain knowledge. The algebraic system
GAP~\cite{GAP4}  has a number of packages for specific types of algebraic
structures. The Small Groups library~\cite{besche-ijac02} contains
\emph{all} ($\approx 4\times 10^8$) non-isomorphic groups up to order 2000 (except for order 1024).
Similarly, the package Smallsemi~\cite{smallsemi} catalogues semigroups and the package LOOPS
catalogues loops~\cite{loops3.4.1}.

Normal forms are ubiquitous in computer science and mathematics, e.g.\
the system \nauty~\cite{McKayP14} uses canonical labeling to decide
isomorphism of graphs.
A large body of research exists on \emph{symmetry breaking} in SAT and
CP~\cite{cp-handbook,sakallah21}.  Computational complexity has been studied
under various notions of lex-leader~\cite{NarodWalshKats,orderings,Luks2004}. We
are not aware, however, of any study of lex-leader in the context of constraints
stemming from first order logic.
\citet{janota-mlex} tackle the calculation of the lex-leader for
\emph{one fixed given} algebra by using SAT\@.

%
Some symmetry breaks are designed to
be fast, when used dynamically, or should add a small number of constraints,
when used statically~\cite{Codish2018,codish-aaai20}.
Such symmetry breaking is often partial such as the least number
heuristic (see Partial Symmetry Breaks section).  \citet{heule-mcs19}
explores optimal complete symmetry breaking for small
graphs~($\approx 5$ vertices) from a theoretical perspective.
\citet{szeider:cp21} develop a specific dynamic symmetry breaking,
called~\emph{SAT Modulo Symmetries}, where a SAT solver is enhanced to
look for the lexicographically smallest graph; similarly \citet{orderly} use \emph{orderly generation}
with the objective to enumerate graphs with certain properties.

For some structures, closed forms are known for calculating the number
of non-isomorphic objects.  For instance, a closed formula is known for
magmas~\cite{harrison1966}, and in the same paper, the author claims that a
closed formula for groups, monoids, and rings might be possible by modifying
the techniques he used, but so far, nobody has managed to find those formulas.

To give an idea of how difficult it is to say something about the size
sequence, we recall two old conjectures: almost all finite groups have size a
power of 2; almost all semigroups are 3-nilpotent (that is, semigroups with
zero, $0$, satisfying the identity $xyz=0$). The solution to the conjecture on
semigroups, widely believed to be true, was announced by~\citet{kleitman1976},
but the proof has a gap that nobody could fix so far.

Given the problems with the closed formula, mathematicians turn to
computational methods to find the first terms of the size sequence.
Traditionally, taking advantage of the deep knowledge of some class to trim the
search tree in a way that usually only works for that class.  Probably the
greatest achievement has been the computation of the number of order 10
semigroups, as the final piece of a long story: in 1955 it was computed up to
size 4~\cite{semigroups4}; in 1977, up to size 7~\cite{juergensen1977}; in
1994, for 8~\cite{satoh1994}; in 2010 for order 9 in Distler's PhD thesis, then
published in journal~\cite{distler2014}; and finally in 2012 for size
10~\cite{distler2012} (by using a combination of non-compact lex-leader
encoding and deep understanding of the problem). Roughly speaking, once a value
was computed, it took about 20 years to get the next one. The OEIS includes
many more size sequences and countless pointers to the bibliography. In
contrast, we introduce a general, out-of-the-box tool that improves upon the
existing methodologies for determining size sequences of algebraic structures.

More broadly, this paper is related to the SAT+CAS program, where SAT is
combined with \emph{computer algebra systems},
cf.~\citet{Williamson,bright-cacm22}.


\section{Conclusions and Future Work}%
\label{sec:conc}

This paper designs a method to calculate a compact complete symmetry break for
finite models of first order logic theory. Such symmetry breaks open an avenue
for SAT-based approaches in computational algebra. We demonstrate the strength
of our approach on the problem of counting  all non-isomorphic algebras of a
fixed size and class. Since we encode into SAT that only canonic (lex-leader)
algebras should be considered, the number of models of the produced SAT formula
is equal to the number of isomorphism classes. Therefore, we can directly apply
model-counting tools to count them \emph{without} enumerating them.
Here we apply exact model counters, but approximate model counting~\cite{amc}
would also be applicable if one only aims for approximate values.

Counting the structures of a given size (\emph{size sequence}
calculation) is an important and old mathematical discipline---it is no
surprise that the first entry of the On-Line Encyclopedia of Integer Sequences
(OEIS) is a size sequence of groups (A000001). Our approach provides a
theory-agnostic tool that further advances this sub-field of universal algebra.

Model counting is not the only possible application enabled by the
complete symmetry break that we calculate. It also enables further SAT-based
reasoning answering further questions, e.g.\  ``Is there an algebra with the property
X?'', ``Do all these algebras have a specific property?''.
The paper also opens several theoretical questions. Namely, what are the
lower/upper bounds for the sizes of canonizing sets for a specific class of
algebras?  Experimentally, the class of magmas requires $n!-1$ permutations
to provide a complete symmetry break, can this
be proven for all domain sizes $n$?
We also intend to extend the tool to support multiple function symbols.


\section*{Acknowledgments}
We thank Chad Brown and Thibault Gauthier for insightful discussions. 
This work was supported by MEYS through the ERC~CZ program under the project \emph{POSTMAN} no.~LL1902, and 
by FCT --- Funda\c{c}\~{a}o para a Ci\^{e}ncia e a Tecnologia, I.P., via the projects UIDB/00297/2020 (\url{doi.org/10.54499/UIDB/00297/2020}) and UIDP/00297/2020 (\url{doi.org/10.54499/UIDP/00297/2020}) (Center for Mathematics and Applications), co-funded by the European Union under the project \emph{ROBOPROX} (reg.~no.~CZ.02.01.01/00/22\_008/0004590).

\bibliography{bibliography}

\end{document}